\documentclass[final,5p,times,twocolumn,sort&compress]{elsarticle}

\usepackage{amssymb}
\usepackage{amsmath}
\usepackage[T1]{fontenc}
\usepackage{lmodern}

\usepackage{bbm}
\usepackage{epsfig}
\usepackage{epstopdf}
\usepackage{color}
\usepackage{float}

\journal{Complexity}
\begin{document}

\begin{frontmatter}



\title{Homogeneous symmetrical threshold model with nonconformity: independence vs. anticonformity.}


\author{Bart{\l}omiej Nowak}
\ead{bartlomiej.nowak@pwr.edu.pl}

\author{Katarzyna Sznajd-Weron}
\ead{katarzyna.weron@pwr.edu.pl}
\address{Department of Theoretical Physics, Faculty of Fundamental Problems of Technology, Wroc\l{}aw University of Science and Technology, Wroc\l{}aw, Poland}

\begin{abstract}
We study two variants of the modified Watts threshold model with a noise (with nonconformity, in the terminology of social psychology) on a complete graph. Within the first version, a noise is introduced via so-called independence, whereas in the second version anticonformity plays the role of a noise, which destroys the order.  The modified Watts threshold model, studied here, is homogeneous and posses an up-down symmetry, which makes it similar to other binary opinion models with a single-flip dynamics, such as the majority-vote and the $q$-voter models. Because within the majority-vote model with independence only continuous phase transitions are observed, whereas within the $q$-voter model with independence also discontinuous phase transitions are possible, we ask the question about the factor, which could be responsible for discontinuity of the order parameter. We investigate the model via the mean-field approach, which gives the exact result in the case of a complete graph, as well as via Monte Carlo simulations. Additionally, we provide a heuristic reasoning, which explains observed phenomena. We show that indeed, if the threshold $r=0.5$, which corresponds to the majority-vote model, an order-disorder transition is continuous. Moreover, results obtained for both versions of the model (one with independence and the second one with anticonformity) give the same results, only rescaled by the factor of 2. However, for $r>0.5$ the jump of the order parameter and the hysteresis is observed for the model with independence, and both versions of the model give qualitatively different results.
\end{abstract}


\end{frontmatter}


\section*{Introduction}
\label{sec:intro}
Models of opinion dynamics are among the most studied models of complex systems \cite{Cas:For:Lor:09,Gal:12,Sen:Cha:13,Sir:Lor:Ser:17}. This is not surprising, because they can be treated as a zero-level approach to various more complex social processes, including polarization of opinion \cite{Def:etal:00,Heg:Kra:02,Kru:Szw:Wer:17}, diffusion of innovation \cite{Szn:etal:14,Byr:etal:16,Wer:Kow:Wer:18} or political voting \cite{Gal:07,Gal:17,Gal:18}.
In most of these models, public opinion is formed as an outcome from individual opinions of mutually interacting agents.  Particularly interesting is a subset of the binary opinion models, including the voter model \cite{Lig:85}, the majority-vote model \cite{Lig:85,Tom:Oli:San:91,Oli:92}, the Galam model \cite{Gal:90}, the Sznajd model  \cite{Szn:Szn:00}, the Watts threshold model \cite{Wat:02}, the $q$-voter model  \cite{Cas:Mun:Pat:09} or the threshold $q$-voter model \cite{Nyc:Szn:13,Vie:Ant:18}. All these models belong to the broader class of binary-state dynamics, likewise the kinetic Ising models \cite{Gle:13}.
The binary decision/opinion framework is not only attractive from physicist's point of view but also natural in the social sciences ~\cite{Gra:78,Wat:02}.

In this paper, we will focus on a particular subclass of the binary opinion models with a single-flip dynamics, which means that one agent at most can change her/his state in a single update \cite{Jed:Szn:18}. Such an updating scheme is used within the voter model, the majority-vote model, the Watts threshold model and the $q$-voter model. There are several common features possessed by all these models, at least in their original formulations. Within these models:
\begin{enumerate}
	\item We consider $N$ individuals  that are tied to the nodes of some graph. Each node of a graph is occupied by exactly one agent.
	\item Each individual is described by the dynamical binary variable $S_{i}(t) = \pm 1, \; i=1, \ldots, N$, that represents an opinion on a given subject (yes/no, agree/disagree, etc.) at given time $t$. Such a variable reminds an Ising spin and therefore wording "individual", "agent", "voter" and "spin" is used interchangeably; $S_i=+1$ is often represented by $\uparrow$, whereas $S_i=-1$ is  by $\downarrow$.
	\item Interactions between agents are local, i.e., they take place only if two agents are directly linked.
	\item At each elementary update a single spin is randomly chosen and it can flip to the opposite direction with a probability that depends on the model's details.
	\item Conformity, i.e. an act of matching opinions, attitudes, beliefs, and/or behaviors to the certain group of influence, is the main type (often the only type) of the social response.
	\item Agents are memoryless, which means that opinion $S_i(t)$ of a given agent at time $t$ depends only on her/his own opinion at the previous time step $S_i(t-\Delta t)$ and opinions of her/his neighbors also at the previous time step $t-\Delta t$.
\end{enumerate}
The differences between models depend mainly on the condition under which the conformity takes place. In some models all neighbors of a given target agent influence her/him (this applies to the majority-vote model \cite{Lig:85,Oli:92} or the Watts threshold model \cite{Wat:02}), whereas in others only a certain group of influence is chosen from the neighborhood (e.g. in the $q$-voter model \cite{Cas:Mun:Pat:09,Nyc:Szn:Cis:12,Nyc:Szn:13}). In the $q$-voter model unanimity of opinion is needed to influence a voter \cite{Cas:Mun:Pat:09}, whereas in the majority-vote model absolute majority is sufficient \cite{Oli:92}. The last and particularly important difference consists of the presence or the absence of the up-down symmetry. The linear voter, the $q$-voter model and majority-vote are symmetrical, i.e., they are invariant to the swap of state labels, as the Ising model without the external field. However, the Watts threshold model is not symmetrical in its original formulation, because it was introduced as a model of innovation diffusion and flips from $\uparrow$ to $\downarrow$ were forbidden \cite{Wat:02,Wat:Dod:07}. Moreover, the Watts threshold model is heterogeneous, even on homogeneous graphs,  i.e. each agent is characterized by its individual threshold needed for conformity. However, modification of the model to make it symmetrical and homogeneous (in a sense that all agents are characterized by the same threshold) is straightforward and will be investigated in this work.

One may ask: what is the motivation to modify the Watts threshold model into the symmetric case? First of all, opinion dynamics concerns not only of asymmetrical problems, as the diffusion of innovations, but many symmetric or almost symmetric issues, such as voting to one of two political parties, choosing one of two products on the duopoly market, etc.  Although the Watts threshold model was originally introduced to model the diffusion of innovation, the main idea of the threshold is also very natural in the broader context. It has been shown in many social experiments that simple majority ($>50\%$) may not be sufficient to convince people, for short review see \cite{Nyc:etal:18}.
The second reason for the modification of the Watts threshold model comes from the basic research. Most of the binary opinion models within sociophysics have the yes-no (up-down) symmetry, so to make a comparison with them it is necessary to deal with the symmetric version of the  Watts threshold model. In fact, the second reason was our main motivation. We wanted to understand the nature of the phase transitions observed within models of binary opinions with a single-flip dynamics and up-down symmetry.

Because all binary models, mentioned above, have been extensively investigated for years, many modifications and extensions of their original formulations have been proposed; a short review on modifications of the majority-vote model can be found in \cite{Vie:Cro:16}, on the Watts threshold model in \cite{Oh:Por:18}, on the $q$-voter model in \cite{Jed:Szn:19} and on the Galam model comprehensive review has been written by the author of the model \cite{Gal:08,Gal:13}. Among many extensions, going into different directions, the introduction of an additional type of the social response was particularly interesting from the point of view of social/psychological sciences, as well as the theory of non-equilibrium phase transitions. This new type of social response is called nonconformity and can take one of two possible forms: (1) independence (resisting influence) or (2) anticonformity (rebelling against influence) \cite{Nai:Mac:Lev:00,Nai:Dom:Mac:13,Nyc:Szn:13}. In the first case, the situation is evaluated independently of the group norm, which means that a state of a given spin is not affected by its neighborhood. In the second case, a voter is influenced by the others but takes the position that is opposite to the group of influence. Therefore it is said that anticonformity and conformity are opposites at the operational level but at the same time similar at the conceptual level, because both indicate behavior that has been influenced by the source \cite{Nai:Dom:Mac:13}.

It is clear that conformity increases agreement (ferromagnetic order) in the system, whereas both types of nonconformity act against consensus. In result of this competition, an order-disorder phase transition emerges. Interestingly, the type of the phase transition (continuous or discontinuous) may depend on the type of nonconformity. For example, it has been shown that within the $q$-voter model with anticonformity only continuous phase transitions are possible, whereas for the $q$-voter model with independence tricriticality (a switch between continuous and discontinuous phase transition) appears \cite{Nyc:Szn:Cis:12,Nyc:Szn:13,Jed:17,Chm:Szn:15,Per:etal:18}. The majority-vote model contains of conformity and anticonformity in its original formulation, but recently an additional noise, in the form of independence, has been introduced \cite{Vie:Cro:16,Enc:etal:19}. It has been shown that the presence of an additional noise does not affect the type of the phase transition, which remains continuous independently of the network structure.

The question that naturally arises here is: "Which factor is responsible for the discontinuous phase transition within the $q$-voter model, since we do not observe the analogous phenomenon within the majority-vote model?" We believe that investigating the modified version of the threshold model on the complete graph could help to understand this phenomenon. Due to our knowledge, the role of a noise has not been explored yet within the Watts threshold model \cite{Wat:02,Oh:Por:18}. Therefore in this paper we will introduce two versions of the model, analogously as it was done for the $q$-voter model, one with independence and the second one with anticonformity.

The paper is organized as follows. In the next section, we describe original versions of binary opinion models with single-flip dynamics. Then we present model's extensions, which consist of introducing the noise into the models and we describe briefly the results that show how this noise impacts phase transitions. In the following subsection, we modify the original Watts threshold model to make it symmetrical and homogeneous, which makes it comparable to other binary opinion models with single-flip dynamics. Subsequently, we propose two versions of the symmetrical, homogeneous Watts threshold model, one with independence and the second one with anticonformity. Then, we analyze the model on the complete graph, which corresponds to the mean-field approach. We compare results obtained within Monte Carlo simulations results with those obtained via analytical treatment. Moreover, we provide a heuristic explanation of the obtained results. Finally, we discuss results in the context of other binary opinion models with single-flip dynamics.

\section*{Methods}
\subsection*{Binary opinion models with a single-flip dynamics}
The general framework of all binary opinion models with a single-flip dynamics has been described above so we will not repeat it here. Instead, we present updating rules that define the dynamics of models within this class. The most extensively studied among all is the linear voter model \cite{Lig:85}. On the other hand, the linear voter model is a special case of the more general  $q$-voter model \cite{Cas:For:Lor:09} and thus we will not discuss it separately. The dynamics of the original $q$-voter model is the following:
\begin{enumerate}
	\item At a given time $t$, choose one voter at random, located at site $i$.
	\item Choose randomly $q$ neighbors of site $i$ from its $k_i$ neighbors, where $k_i$ is degree of a node $i$. In the original formulation and in some later versions repetitions were allowed to make the model universal (arbitrary value of $q$ on the arbitrary graph is possible) \cite{Cas:Mun:Pat:09,Mob:15,Mel:Mob:Zia:16,Mel:Mob:Zia:17}. However, in many other papers repetitions were forbidden \cite{Nyc:Szn:Cis:12,Nyc:Szn:13,Jed:17,Nyc:etal:18}.
	\item If all $q$ neighbors have the same opinion, the spin at site $i$ takes the same state as $q$ neighbors.
	\item Otherwise, i.e. in lack of unanimity, spin at site $i$ can flip to the opposite direction with probability $\epsilon$. In most of later modifications $\epsilon=0$ \cite{Mob:15,Mel:Mob:Zia:16,Mel:Mob:Zia:17,Nyc:Szn:Cis:12,Nyc:Szn:13,Jed:17,Nyc:etal:18} and here we also refer to this case.
	\item Time is updated $t = t + \frac{1}{N}$.
\end{enumerate}
In \cite{Nyc:Cis:Szn:12} two types of noise (interpreted as nonconformity) have been introduced to the model, but not simultaneously. Initially two versions of the model have been introduced: one with independence and the second one with anticonformity. In each of these models nonconformity (independence or anticonformity) takes place with probability $p$, whereas with complementary probability $1-p$ agent conforms. Summarizing, the algorithm of a single step is the following:
\begin{enumerate}
	\item At a given time $t$, choose one voter at random, located at site $i$.
	\item Update the opinion $S_{i}$:
	\begin{itemize}
		\item Model \textbf{I} (with \textbf{I}ndependence)
		\begin{enumerate}
			\item With probability $p$, an agent changes opinion independently, i.e. she/he changes opinion to the opposite one $S_i \rightarrow -S_i$ with probability $\frac{1}{2}$.
			\item With probability $1-p$, an agent conforms, i.e. if all $q$ agents, randomly chosen from all $k_i$ neighbors of site $i$, are in the same state then the voter at site $i$ takes the same position as those $q$ agents.
		\end{enumerate}
		\item Model \textbf{A} (with \textbf{A}nticonformity)
		\begin{enumerate}
			\item With probability $p$, an agent anticonforms, i.e. acts against a group of influence, i.e. if all $q$ agents, randomly chosen from all  $k_i$ neighbors of site $i$, are in the same state then the voter at site $i$ takes the opposite position to those $q$ agents.
			\item With probability $1-p$, an agent changes opinion as in the Model I.
		\end{enumerate}
	\end{itemize}
	\item Time is updated $t = t + \frac{1}{N}$.
\end{enumerate}
The generalized versions of the model that consists of a threshold \cite{Nyc:Szn:13,Vie:Ant:18} and two types of nonconformity simultaneously \cite{Nyc:etal:18} has been also introduced. In the case of the threshold $q$-voter model, only $r$ among $q$ neighbors have to share the same opinion in order to influence a voter.

The $q$-voter model with independence was studied on the complete graph  \cite{Nyc:Cis:Szn:12,Nyc:Szn:13,Nyc:etal:18}, as well as on various complex networks \cite{Jed:17,Chm:Szn:15}, whereas the $q$-voter model with anticonformity only on the complete graph. It has been shown, that within Model A only continuous phase transitions are possible for all $q \ge 2$, whereas within Model I both types of phase transitions appear: for $2 \le q \le 5$ there is continuous phase transition and for $q>5$ transition is discontinuous. It has occurred that the tricritical point $q^*=5$, even if $q$ can take non-integer values \cite{Per:etal:18}.

It has been also shown that in the case of the threshold $q$-voter model there is a critical threshold $r^*=r^*(q)$ that decreases with $q$, above which discontinuous phase transitions are possible \cite{Nyc:Szn:13,Nyc:etal:18}. In the most general case, when independence and anticonformity are introduced simultaneously, it occurs that $r^*=r^*(q)$ is monotonically decreasing function of $q$ and $r^*$ is always greater than $0.5$, even for very large $q$ \cite{Nyc:etal:18}. Although, the analytical form of $r^*=r^*(q)$ has not been found, it was predicted that for $q \rightarrow N$ the critical value $r^* \rightarrow 0.5$. It means that the absolute majority is not sufficient for the discontinuous phase transition, even if for very large $q$.

Another model with a single-flip dynamics, that has been analyzed in the presence of a noise, is the majority-vote model \cite{Lig:85}. The dynamics of the original majority-vote model is the following:
\begin{enumerate}
	\item At a given time $t$, choose one voter at random, located at site $i$.
	\item With probability $f$ a spin at site $i$ adopts the minority sign of  $k_i$ neighboring spins.
	\item With complementary probability $1-f$ a spin at site $i$ adopts the majority sign of  $k_i$ neighboring spins.
	\item Time is updated $t = t + \frac{1}{N}$.
\end{enumerate}
As seen from the above description, within the original majority-vote model anticonformity takes place with probability $f$ and conformity with $1-f$, similarly as within the $q$-voter model with anticonformity. Recently an additional noise has been introduced into the majority-vote model: with probability $p$ a voter acts independently, i.e., analogously like within the $q$-voter model with independence, flips randomly to the opposite direction with probability 1/2. With complementary probability $1-p$ the original rule is applied \cite{Vie:Cro:16,Enc:etal:19}. The model has been investigated on the square lattice \cite{Vie:Cro:16}, as well as on several graphs, including homogeneous and heterogeneous structures, through the mean-field calculations and the Monte Carlo simulations \cite{Enc:etal:19}. It has been shown that only continuous phase transitions are possible within this model, similarly as for the model without additional noise. Discontinuous phase transitions do not appear even for highly connected networks. This result is consistent with the result obtained for the generalized threshold $q$-voter model \cite{Nyc:Szn:13,Nyc:etal:18}, as described above.

\subsection*{Symmetrical, homogeneous Watts threshold model}
\label{sec:model}
In computational sociology, a particularly popular class of models describing the spread of innovation/idea/behavior are threshold models, based on the idea introduced by Granovetter \cite{Gra:78}. Probably the simplest among them is the Watts threshold model \cite{Wat:02}. The updating rule within its original formulation is the following:
\begin{enumerate}
	\item At a given time $t$, choose one voter at random, located at site $i$.
	\item An agent at site $i$ is influenced by its $k_i$ neighbors. If at least a threshold fraction $r_i$ of its $k_i$ neighbors are in state $1$ then an agent adopts this state, otherwise nothing happens.
	\item Time is updated $t = t + \frac{1}{N}$.
\end{enumerate}
There are two characteristic features that make the model different from other models described in the previous subsection. The first visible difference is heterogeneity. In other models it is introduced only by the heterogeneity of a graph, here agents posses individual thresholds. Originally, each agent is assigned a threshold that is drawn at random from a given probability distribution function (PDF). Of course, as a special case, we can choose a one-point PDF, which takes the value equal to one at $r$, and zero otherwise. Within such a formulation $r$ is an external (control) parameter of the model.

Another difference between Watts threshold model and other models, presented above, is the lack of the up-down symmetry. An agent who ones adopted cannot go back to an unadopted state. However, we can easily modify the model to make it symmetrical, in the following way:
\begin{enumerate}
	\item At a given time $t$, choose one voter at random, located at site $i$.
	\item An agent at site $i$ is influenced by all  $k_i$ neighbors:
	\begin{enumerate}
		\item if at least a threshold fraction $r$ of its $k_i$ neighbors are in state $1$ then an agent takes state 1, else
		\item if at least a threshold fraction $r$ of its $k_i$ neighbors are in state $-1$ then an agent takes state -1, else
		\item an agent remains in its old state.
	\end{enumerate}
	\item Time is updated $t = t + \frac{1}{N}$.
\end{enumerate}

It should be noticed that within the above definition, the rules are defined unambiguously only for  $r \in [0.5,1]$. To clarify this, let us give here an example. Imagine that $r=0.3$ and at a given time step $c(t)=0.4$. It means that the ratio of the positive opinions is equal to $0.4$ and the ratio of the negative opinions is equal to $0.6$. Because $r=0.3$ both conditions: (a) a threshold fraction $r$ of neighbors are in state $1$, as well as (b) a threshold fraction $r$ of neighbors are in the state $-1$ are fulfilled. Which means that there is no unambiguous choice. Of course one could think about another model in which this ambiguity could be solved by introducing the probabilistic rule. However, in such a case interpretation of what is conformity and what is anticonformity would be far less clear. Therefore here we consider only $r \in [0.5,1]$. This can be interpreted also as a supermajority (or a qualified majority), whereas $r=0.5$ corresponds to the simple majority rule.

Now we are ready to introduce two versions of the model with nonconformity, one with \textbf{I}ndependence (Model \textbf{I}) and the second one with \textbf{A}nticonformity (Model \textbf{A}), analogously as it was done for $q$-voter model.
Within Model A an agent conforms or anticonforms if the fraction of neighboring spins having the same state is larger than a fixed threshold $r$. In the conformity case (which takes place with probability $1-p$), an agent follows the opinion of the group of influence, whereas in the case of anticonformity (which takes place with probability $p$) she/he takes the opposite opinion to the group, as in the majority-vote or in the $q$-voter model with anticonformity. Within Model I, instead of anticonformity,  independence takes place with probability $p$: an agent flips to the opposite state with probability $1/2$.

The algorithm of a single update is the following:
\begin{enumerate}
	\item At a given time $t$, choose one voter at random, located at site $i$.
	\item Update the opinion $S_{i}$:
	\begin{itemize}
		\item Model I (with Independence)
		\begin{enumerate}
			\item With probability $p$, an agent acts independently, i.e. she/he changes an opinion to the opposite one $S_i \rightarrow -S_i$ with probability $\frac{1}{2}$.
			\item With probability $1-p$, an agent conforms to its  $k_i$ neighbors:
			\begin{enumerate}
			\item if at least a threshold fraction $r$ of its $k_i$ neighbors are in state $1$ then an agent takes state $1$, else
			\item if at least a threshold fraction $r$ of its $k_i$ neighbors are in state $-1$ then an agent takes state $-1$, else
			\item an agent remains in its old state.
			\end{enumerate}
		\end{enumerate}
		\item Model A (with Anticonformity)
		\begin{enumerate}
			\item With probability $p$, an agent anticonforms to its  $k_i$ neighbors:
			\begin{enumerate}
			\item if at least a threshold fraction $r$ of its $k_i$ neighbors are in state $1$ then an agent takes state $-1$, else
			\item if at least a threshold fraction $r$ of its $k_i$ neighbors are in state $-1$ then an agent takes state $1$, else
			\item an agent remains in its old state.
			\end{enumerate}
			\item With probability $1-p$, an agent conforms ot its neighbors, analogously as in Model I.
		\end{enumerate}
	\end{itemize}
	\item Time is updated $t = t + \frac{1}{N}$.
\end{enumerate}

\subsection*{The mean-field approach}
In this work, analogously as in \cite{Nyc:Szn:Cis:12}, we analyze the model on a complete graph, which means that for each agent, all other agents in the system are neighbors. On one hand side, one can argue that in the case of the symmetric homogeneous threshold model such a structure will give trivial, easily predictable results. On the other side, only for the complete graph the mean-field approach is exact. Moreover, most of the results for $q$-voter model with nonconformity were obtained on the complete graph, so this structure is adequate for comparison between models. Finally, we hope that such an approach will be helpful in understanding the nature of the phase transitions within the $q$-voter model and the majority-vote model, what will be discussed in the last section of this paper.

As an aggregated quantity, which fully describes the system in case of the complete graph, we choose an average concentration of agents with positive opinions:
\begin{equation}
c(t) = \frac{N_{\uparrow}(t)}{N},
\end{equation}
where $N_{\uparrow}(t)$ denotes the number of agents in the state $\uparrow$ at time $t$. Alternatively, we could choose an average opinion (magnetization), which is a natural order parameter \cite{Lan:37}:
\begin{equation}
m(t) = \frac{1}{N} \sum_{i=1}^N S_i(t)=\frac{N_{\uparrow}(t)-N_{\downarrow}(t)}{N}=2c(t)-1.
\end{equation}
However, for the analytical treatment $c(t)$ is more convenient, because within the mean-field approach it gives the probability that a randomly chosen agent, at any site, is positive.

As for all other binary opinion models with a single-flip dynamics, in an elementary time step the number of up spins $N_{\uparrow}$ can increase by one, decrease by one or remain the same. It means that the concentration of up spins $c$ can change only by $\pm \frac{1}{N}$. We follow the notation from \cite{Nyc:Szn:Cis:12,Nyc:Szn:13}:
 \begin{equation}
\begin{aligned}
\gamma^{+} &= Prob\left(c (t + \Delta t) = c(t) + \frac{1}{N}\right), \\
\gamma^{-} &= Prob\left(c (t + \Delta t) = c(t) - \frac{1}{N}\right),
\end{aligned}
\end{equation}
where $\Delta t = \frac{1}{N}$, as usually. Of course with the complementary probability $1-\gamma^{+}-\gamma^{-}$ the state of the system will not change.

Using above probabilities, we obtain a recursive formula for the concentration of up spins:
\begin{equation}
c \left(t + \frac{1}{N} \right)=c(t)+\frac{1}{N}\left(\gamma^+ -\gamma^-\right),
\label{eq:c_iter}
\end{equation}
which for $N\rightarrow\infty$ gives the rate equation \cite{Hin:00,Kra:Red:Ben:10}:
\begin{equation}
\frac{dc(t)}{dt}=\gamma^+ -\gamma^-.
\label{eq:rate}
\end{equation}

Probabilities $\gamma^+, \gamma^-$ depend of course on model's details and can be derived looking at all possible changes that may occur in a single update. Within the homogeneous symmetrical threshold model with independence (Model I) the following changes are possible:
\begin{equation}
\begin{split}
        \underbrace{\uparrow \uparrow \dots \uparrow}_{\geqslant \lfloor{r(N-1)}\rfloor} \Downarrow &\stackrel{1-p}{\longrightarrow} \underbrace{\uparrow \uparrow \dots \uparrow}_{\geqslant \lfloor{r(N-1)}\rfloor} \Uparrow, \\
        \underbrace{\downarrow \downarrow \dots \downarrow}_{\geqslant \lfloor{r(N-1)}\rfloor} \Uparrow &\stackrel{1 - p}{\longrightarrow} \underbrace{\downarrow \downarrow \dots \downarrow}_{\geqslant \lfloor{r(N-1)}\rfloor} \Downarrow ,\\
        \underbrace{\dots\dots\dots}_{\substack{\text{any} \\ \text{configuration}}} \Uparrow &\stackrel{p/2}{\longrightarrow} \underbrace{\dots\dots\dots}_{\substack{\text{any} \\ \text{configuration}}} \Downarrow, \\
        \underbrace{\dots\dots\dots}_{\substack{\text{any} \\ \text{configuration}}} \Downarrow &\stackrel{p/2}{\longrightarrow} \underbrace{\dots\dots\dots}_{\substack{\text{any} \\ \text{configuration}}} \Uparrow, \\
\end{split}
\end{equation}
where $\Downarrow$ and $\Uparrow$ denotes states of a target agent, and $\lfloor{r(N-1)}\rfloor$ is the floor function of $r(N-1)$, which follows from the restriction that the number of agents can take only integer value, whereas $r$ is a real number.

Within the homogeneous symmetrical threshold model with anticoformity (Model A):
\begin{equation}
\begin{split}
        \underbrace{\uparrow \uparrow \dots \uparrow}_{\geqslant \lfloor{r(N-1)}\rfloor} \Downarrow &\stackrel{1 - p}{\longrightarrow} \underbrace{\uparrow \uparrow \dots \uparrow}_{\geqslant \lfloor{r(N-1)}\rfloor} \Uparrow \\
        \underbrace{\downarrow \downarrow \dots \downarrow}_{\geqslant \lfloor{r(N-1)}\rfloor} \Uparrow &\stackrel{1-p}{\longrightarrow} \underbrace{\downarrow \downarrow \dots \downarrow}_{\geqslant \lfloor{r(N-1)}\rfloor} \Downarrow \\
        \underbrace{\uparrow \uparrow \dots \uparrow}_{\geqslant \lfloor{r(N-1)}\rfloor} \Uparrow &\stackrel{p}{\longrightarrow} \underbrace{\uparrow \uparrow \dots \uparrow}_{\geqslant \lfloor{r(N-1)}\rfloor} \Downarrow \\
        \underbrace{\downarrow \downarrow \dots \downarrow}_{\geqslant \lfloor{r(N-1)}\rfloor} \Downarrow &\stackrel{p}{\longrightarrow} \underbrace{\downarrow \downarrow \dots \downarrow}_{\geqslant \lfloor{r(N-1)}\rfloor} \Uparrow \\
\end{split}
\end{equation}
In all other situations, the state of the system will not change. Therefore:
\begin{equation}
\begin{gathered}
\gamma^{+} = (1-p)\alpha_{\uparrow} + p\beta_{\uparrow} ,\\
\gamma^{-} = (1-p)\alpha_{\downarrow} + p\beta_{\downarrow},
\end{gathered}
\end{equation}
where $\alpha_{\uparrow},\alpha_{\downarrow}$ are probabilities related to conformity. They are the same for both versions of the model:
\begin{equation}
\begin{split}
\alpha_{\uparrow} &= \sum_{i = \lfloor{r(N-1)}\rfloor}^{N-1} \left[\rule{0cm}{1cm}\right.\binom{N-1}{i} \times \\
&\times \frac{N_{\downarrow} \prod\limits_{j=1}^{i} \left(N _{\uparrow}-j+1\right)\prod\limits_{j=1}^{N-1-i} \left(N_{ \downarrow }-j+1\right)} {\prod\limits_{j=1}^{N}( N - j + 1 )}\left.\rule{0cm}{1cm}\right] \\
\alpha_{\downarrow} &= \sum_{i = \lfloor{r(N-1)}\rfloor}^{N-1}\left[\rule{0cm}{1cm}\right.\binom{N-1}{i}  \times \\ &\times\frac{N_{\uparrow} \prod\limits_{j=1}^{i} \left(N _{\downarrow}-j+1\right)\prod\limits_{j=1}^{N-1-i} \left(N_{ \uparrow }-j+1\right)} {\prod\limits_{j=1}^{N}( N - j + 1 )}\left.\rule{0cm}{1cm}\right].
\end{split}
\end{equation}
Within conformity, an agent accepts the opinion of her/his neighbors, i.e. it takes the same state as the majority above the threshold $r$. Thus the probability $\alpha_{\uparrow}$ that an agent changes her/his opinion from $-1$ to $1$ is equal to the probability that a randomly selected agent is at state $-1$ multiplied by the probability that at least $\lfloor{r(N-1)}\rfloor$ of her/his neighbors are in the state $1$. Analogously, the probability $\alpha_{\downarrow}$ that an agent changes her/his opinion from $1$ to $-1$ is equal to the probability that a randomly selected agent is at state $1$ multiplied by the probability that at least $\lfloor{r(N-1)}\rfloor$ of her/his neighbors are in the state $-1$.

On the other hand, $\beta_{\uparrow}$, $\beta_{\downarrow}$ are related to nonconformity, i.e. They depend on the model's version. In the model with independence an agent changes her/his state to the opposite one with probability $\frac{1}{2}$. Thus $\beta_{\uparrow}$ is equal to the probability that a randomly selected agent is at state $-1$ multiplied by $1/2$. Analogously, $\beta_{\downarrow}$ is equal to the probability that a randomly selected agent is at state $1$ multiplied by $1/2$ and thus:
\begin{equation}
\begin{gathered}
\beta_{\uparrow} = \frac{N_{\downarrow}}{2N},\\
\beta_{\downarrow} = \frac{N_{\uparrow}}{2N}.
\end{gathered}
\end{equation}
Whereas for anticonformity:
\begin{equation}
\begin{split}
\beta_{\uparrow} &= \sum_{i = \lfloor{r(N-1)}\rfloor}^{N-1} \left[\rule{0cm}{1cm}\right.\binom{N-1}{i} \times \\
&\times \frac{N_{\downarrow} \prod\limits_{j=1}^{i} \left(N _{\downarrow}-j+1\right)\prod\limits_{j=1}^{N-1-i} \left(N_{ \uparrow }-j+1\right)} {\prod\limits_{j=1}^{N}( N - j + 1 )}\left.\rule{0cm}{1cm}\right] \\
\beta_{\downarrow} &= \sum_{i = \lfloor{r(N-1)}\rfloor}^{N-1}\left[\rule{0cm}{1cm}\right.\binom{N-1}{i}  \times \\ &\times\frac{N_{\uparrow} \prod\limits_{j=1}^{i} \left(N _{\uparrow}-j+1\right)\prod\limits_{j=1}^{N-1-i} \left(N_{ \downarrow }-j+1\right)} {\prod\limits_{j=1}^{N}( N - j + 1 )}\left.\rule{0cm}{1cm}\right].
\end{split}
\end{equation}
Within anticonformity, an agent takes the opposite state to her/his neighbors, which are in the majority above the threshold $r$. Thus the probability $\beta_{\uparrow}$ that an agent changes her/his opinion from $-1$ to $1$ is equal to the probability that a randomly selected agent is at state $-1$ multiplied by the probability that at least $\lfloor{r(N-1)}\rfloor$ of her/his neighbors are in the state $-1$. Analogously, the probability $\beta_{\downarrow}$ that an agent changes her/his opinion from $1$ to $-1$ is equal to the probability that a randomly selected agent is at state $1$ multiplied by the probability that at least $\lfloor{r(N-1)}\rfloor$ of her/his neighbors are in the state $1$.

For the large systems $N>>1$ we use the following approximation, analogously as in \cite{Nyc:Szn:13}:
\[
\frac{N_{\uparrow} + h}{N + g} \approx c, \quad \frac{N_{\downarrow} + h}{N + g} \approx 1 - c,
\]
where $h,g$ are positive, finite constants.
In such a case equations for  $\gamma^{+}$ and $\gamma^{-}$ take much simpler forms.
For Model I:
\begin{equation}
\begin{split}
\gamma^{+} &= \frac{p(1-c)}{2} + (1-p)\sum_{i = \lfloor{r(N-1)}\rfloor}^{N-1} \binom{N-1}{i} c^{i} (1-c)^{N-i},\\
\gamma^{-} &= \frac{pc}{2} + (1-p)\sum_{i = \lfloor{r(N-1)}\rfloor}^{N-1} \binom{N-1}{i} (1-c)^{i} c^{N-i}.
\end{split}
\label{eq:12}
\end{equation}
For Model A:
\begin{equation}
\begin{split}
\gamma^{+} &= p\sum_{i = \lfloor{r(N-1)}\rfloor}^{N-1} \binom{N-1}{i} (1-c)^{i+1} c^{N-1-i} +\\
&+ (1-p)\sum_{i = \lfloor{r(N-1)}\rfloor}^{N-1} \binom{N-1}{i} c^{i} (1-c)^{N-i},\\
\gamma^{-} &= p\sum_{i = \lfloor{r(N-1)}\rfloor}^{N-1} \binom{N-1}{i} c^{i+1} (1-c)^{N-1-i} +\\
&+ (1-p)\sum_{i = \lfloor{r(N-1)}\rfloor}^{N-1} \binom{N-1}{i} (1-c)^{i} c^{N-i}.
\end{split}
\label{eq:13}
\end{equation}
Summations in above formulas can be calculated using the cumulative distribution function of the binomial distribution and thus we obtain for Model I:
\begin{equation}
\begin{split}
\gamma^{+} &= (1-c)\left[\frac{p}{2} + (1-p) B_{c}\right],\\
\gamma^{-} &= c\left[\frac{p}{2} + (1-p) B_{1-c}\right],\\
\end{split}
\label{eq:gamma_inf_I}
\end{equation}
and for Model A:
\begin{equation}
\begin{split}
\gamma^{+} &= \mathbin{(1-c)}\left[p B_{1-c} + (1-p) B_{c}\right],\\
\gamma^{-} &= c\left[p B_{c} + (1-p) B_{1-c}\right].
\end{split}
\label{eq:gamma_inf_A}
\end{equation}
We use the following notation:
\begin{equation}
\begin{split}
B_{c} &= P(X_{1}\geqslant \lfloor{r(N-1)}\rfloor), \\
B_{1-c} &= P(X_{2}\geqslant~\lfloor{r(N-1)}\rfloor),
\end{split}
\end{equation}
where $X_{1}$ is a binomially distributed random variable with $N-1$ number of trials and success probability in each trial equal to $c$,  and $X_{2}$ is a binomially distributed random variable with $N-1$ number of trials and success probability in each trial $1 - c$:
\begin{equation}
X_{1} \sim B(N-1,c), \quad X_{2} \sim B(N-1,1-c).
\end{equation}
For the system size $N$ large enough, we can approximate binomial distribution by the normal one, which should simplify  calculations. Unfortunately, even within such an approximation, we are not able to  derive an analytical formula for $c(t)$, described by the rate equation (\ref{eq:rate}). However, we can solve the equation numerically or, in a case of the finite system, calculate $c(t)$ by iterating Eq. (\ref{eq:c_iter}).

However, usually, we are more interested in the stationary state than in the time evolution. Especially, that the aim of this work is to understand the nature of phase transitions induced by the noise and thus we are interested in the dependence between the stationary value of the concentration of up spins $c_{st}$ and the level of noise (probability of nonconformity) $p$.

\begin{figure}[h]
	\vskip 0.3cm
	\centerline{\epsfig{file=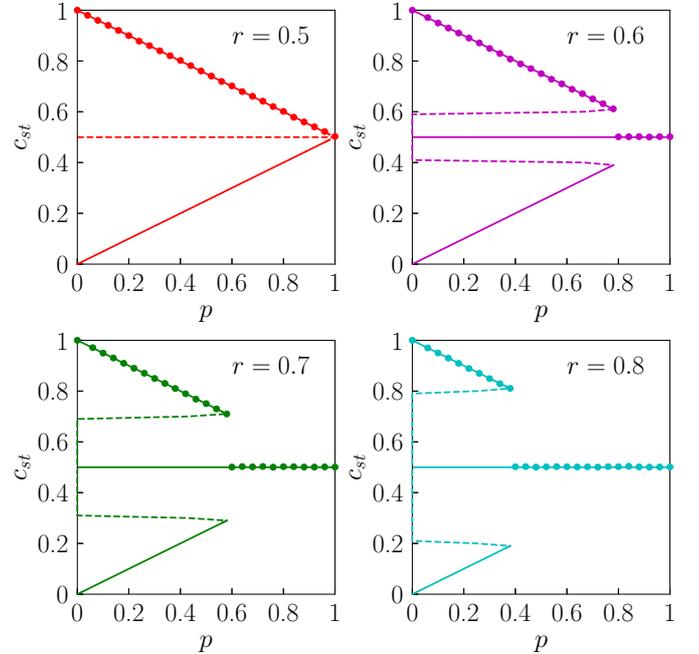,width=1\columnwidth}}
	\caption{Phase diagrams for the model with independence for different values of the threshold $r$. Lines indicate the analytical prediction from MFA and dots represent results of the Monte Carlo simulations from the initial fully ordered state ($c(0)=1$) for the system of size $N = 5 \cdot 10^{4}$.}
	\label{fig:indep_st}
\end{figure}

\begin{figure}[h]
	\vskip 0.3cm
	\centerline{\epsfig{file=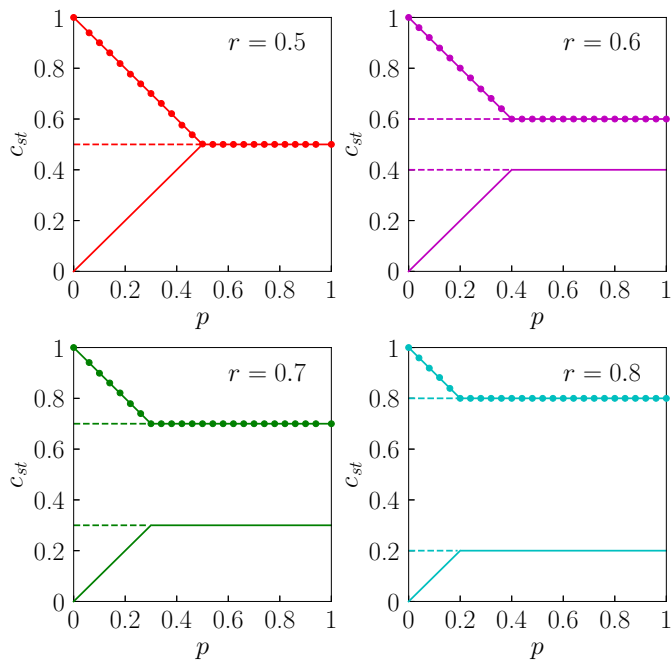,width=1\columnwidth}}
	\caption{Phase diagrams for the model with anticonformity for different values of the threshold $r$. Lines indicate the analytical prediction from MFA and dots represent results of the Monte Carlo simulations from the initial fully ordered state ($c(0)=1$) for the system of size $N = 5 \cdot 10^{4}$.}
	\label{fig:anti_st}
\end{figure}

From Eq. (\ref{eq:c_iter}) we see that in the stationary state, the probability $\gamma_+$ and $\gamma_-$ should be equal. Thus to calculate stationary values of concentration we should simply solve the equation
\begin{equation}
\gamma_{+} - \gamma_{-} = 0.
\label{eq:17}
\end{equation}
Solving analytically Eq. (\ref{eq:17}), i.e., finding $c_{st}$ as a function of $p$ is impossible, but we can easily derive the opposite relations satisfying Eq. (\ref{eq:17}), analogously as it was done for the $q$-voter model \cite{Nyc:Szn:Cis:12}. If we use formulas (\ref{eq:gamma_inf_I}) and  (\ref{eq:gamma_inf_A}) then we obtain for Model I:
\begin{equation}
 p = \frac{c_{st}B_{1-c_{st}} - (1-c_{st})B_{c_{st}}}{\frac{1}{2} - c_{st} - (1-c_{st})B_{c_{st}} + c_{st}B_{1-c_{st}}},
\label{eq:pc_st_I}
\end{equation}
whereas for Model A:
\begin{equation}
p = \frac{B_{c_{st}} - c_{st}(B_{c_{st}} + B_{1-c_{st}})}{B_{c_{st}} - B_{1-c_{st}}}.
    \label{eq:pc_st_A}
\end{equation}
We have used above formulas to plot the dependency between $c_{st}$ and probability $p$ for several values of the threshold $r$ on Figs. \ref{fig:indep_st} and \ref{fig:anti_st}. Although the relation $c_{st}(p)$ is unknown, only the relation $p(c_{st})$ was calculated, we can plot diagrams by rotating the figure, analogously as in \cite{Nyc:Szn:Cis:12}.

\section*{Results}
We have investigated the model within the mean-field approach, described in the previous section, as well as via the Monte Carlo simulations. We have conducted simulations for several system sizes varying from $N=10^3$ to $N=10^5$ and for $N=5 \cdot 10^4$ we have obtained satisfying agreement with formulas obtained within MFA for the large system, what can be seen in Figs. \ref{fig:indep_st} and \ref{fig:anti_st}. Results were averaged only over $10$ samples but this was enough for this size of the system to get the good statistics -- error bars in Figs. \ref{fig:indep_st} and \ref{fig:anti_st} are of the same size or even smaller than the symbols. Solid lines indicate
stable (attracting) steady values of concentration, whereas dashed lines indicate unstable (repelling) steady states.

It is seen that generally, the dependence between $c_{st}$ and $p$ is quite trivial (linear) for both variants of the model. For $r=0.5$, which corresponds to the simple majority, results for Model I and Model A are almost identical, only rescaled. For both models:
\begin{eqnarray}
\begin{array}{l}
c_{st}  =
\left\{
\begin{array}{ll}
1-\lambda p & \mbox{for} \hspace{0.2cm} p \le 1/2\lambda \hspace{0.2cm} \mbox{and}  \hspace{0.2cm}c(0)>1/2 \\
3 \lambda p-1 & \mbox{for} \hspace{0.2cm} p \le 1/2\lambda \hspace{0.2cm} \mbox{and}  \hspace{0.2cm} c(0)<1/2 \\
\frac{1}{2} & \mbox{for} \hspace{0.2cm} p > 1/2\lambda\\
\end{array}
\right.
\end{array}
\label{eq:cst_r05}
\end{eqnarray}
where $\lambda=1/2$ for Model I, whereas $\lambda=1$ for model A. It means that for $p \le p^*=1/2\lambda$ there are two stable solutions: for any initial value $c(0)>1/2$ the system eventually reaches $c_{st}=1-\lambda p$, whereas for any initial value $c(0)<1/2$ the system eventually reaches $c_{st}=3 \lambda p-1$. For $p>p^*=1/2 \lambda$ there is only one stable solution with up-down symmetry, i.e. $c_{st}=1/2$. This reminds a  continuous phase transition, only the dependence $c_{st}(p)$ is trivial.

\textbf{For any threshold $r>0.5$} the situation is slightly more complicated. We can still identify a value $p=p^*$ below which $c_{st}$ decays or increases monotonically with $p$, depending on the initial condition $c(0)$. However, this time it is not reached from arbitrary value of $c(0) \ne 1/2$. In order to reach one of the ordered state, denoted by the solid lines in Figs.~\ref{fig:indep_st}-\ref{fig:anti_st}, the initial concentration of up spins $c(0)$ or alternatively initial concentration of down spins $1-c(0)$ has to be larger than a threshold $r$, i.e. $c(0)>r$ or $1-c(0)>r \rightarrow c(0)<1-r$. Having this in mind, for $p<p^*(r)$ we can actually rewrite Eq. (\ref{eq:cst_r05}):
\begin{eqnarray}
\begin{array}{l}
c_{st}  =
\left\{
\begin{array}{ll}
1-\lambda p & \mbox{for} \hspace{0.2cm} c(0)>r \\
3 \lambda p-1 & \mbox{for} \hspace{0.2cm}  c(0)<1-r, \\
\end{array}
\right.
\end{array}
\label{eq:cst_r}
\end{eqnarray}
and again  $\lambda=1/2$ for Model I, whereas $\lambda=1$ for model A for arbitrary value of $r$.
The threshold $p^*(r)$ can be easily derived from the condition:
\begin{equation}
1- \lambda p^*(r)=r \rightarrow p^*(r)=\frac{1-r}{\lambda}.
\label{eq:p_crit}
\end{equation}
We see that for $r=1/2$, we obtain $p^*=1/2 \lambda$, as expected, and Eq. (\ref{eq:cst_r}) becomes almost identical as Eq. (\ref{eq:cst_r05}).

\begin{figure}[h]
	\vskip 0.3cm
	\centerline{\epsfig{file=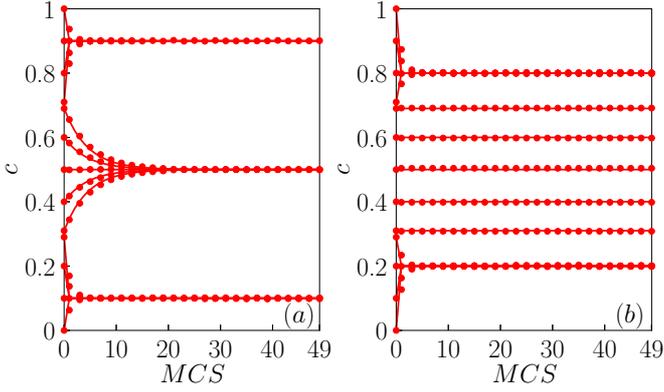,width=1\columnwidth}}
	\caption{Average trajectories for the system of the size $N = 5 \cdot 10^{4}$ for the probability of nonconformity $p=0.2$ and models with $(a)$ independence, $(b)$ anticonformity both with $r = 0.7$ Dots represent an outcome of the Monte Carlo simulations and solid lines refer to MFA results.}
	\label{fig:traject}
\end{figure}

\begin{figure}[h]
	\vskip 0.3cm
	\centerline{\epsfig{file=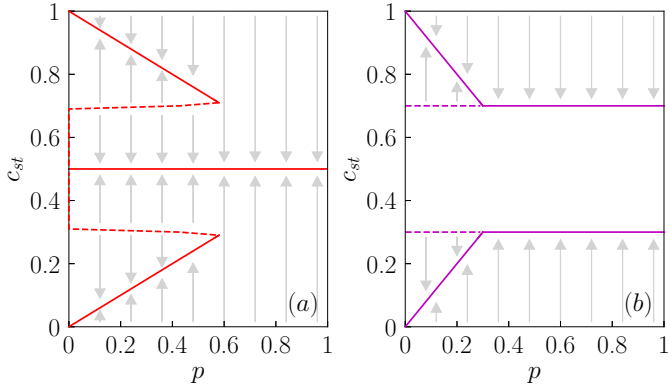,width=1\columnwidth}}
	\caption{Flow diagrams for the threshold model with $r=0.7$ with $(a)$ independence and $(b)$ anticonformity. Solid lines indicates stable steady states, whereas dashed lines represent unstable ones. Arrows indicate the direction of flow in the space of parameters $(c(0),p)$.}
	\label{fig:flow}
\end{figure}

However, there is one crucial difference between the case $r=1/2$ and $r>1/2$. For $r=1/2$, independently on the version of the model, disordered steady state $c_{st}=1/2$ is unstable for $p<p^*$ and stable for $p>p^*$. For $r>0.5$ each version of the model behaves differently. To better illustrate differences between model I and model A we present trajectories for $p<p^*$ on Fig. \ref{fig:traject}, as well as the flow diagrams on Fig. \ref{fig:flow} for a fixed value of $r=0.7$.

\textbf{Let us first discuss results for Model I}, shown in Fig. \ref{fig:indep_st} and on the left panels of Figs.~\ref{fig:traject} and \ref{fig:flow}. In this case, $c_{st}=1/2$ is stable for any value of $p$ and for $p<p^*$ two additional symmetrical unstable steady states appear: $c_{st}=r$ and $c_{st}=1-r$. This means that from the initial state $c(0) \in (1-r,r)$ the system will be attracted to the disordered state for any value of $p$. In result there is jump of size $r-1/2$ at $p=p^*(r)$, where from Eq. (\ref{eq:p_crit}):
\begin{equation}
p^*(r)=2(1-r).
\label{eq:p_crit_MI}
\end{equation}
Moreover, one could also identify hysteresis, because for $p \in (0, 2(1-r))$ the stationary concentration of up spins $c_{st}$ depends on the initial state $c(0)$. So, looking naively at Fig.~\ref{fig:indep_st} one could argue that for the model with independence for any value of $r>1/2$ there is a discontinuous phase transition: the jump of the order parameter increases, whereas hysteresis decreases with growing $r$. Of course, again this result is trivial after a short moment of reflection but we will discuss it later.

\textbf{For the model A}, as shown in Fig.~\ref{fig:anti_st} and on the right panels of Figs.~\ref{fig:traject} and \ref{fig:flow}, there is no jump and even for $p>p^*=1-r$ the system remains ordered, i.e. $c_{st}=r$ if we start from the initial condition $c(0)>r$ or $c_{st}=1-r$ if we start from the initial condition $c(0)<1-r$. For $c(0) \in [1-r,r]$ system does not evolve and $c_{st}=c(0)$.

Probably, some of the readers wonder why we use such a formal approach, which eventually led to Eqs. (\ref{eq:pc_st_I})-(\ref{eq:pc_st_A}), as well as the time-consuming  Monte Carlo simulations, if nearly all results could be easily deduced directly from the model's assumptions, without any calculations. Indeed, in the case of this model it would be possible and will be discussed in the next subsection. However, usually the standard approach that we have presented above is the only possible way to obtain results. That was the case for $q$-voter model with independence for which the Monte Carlo simulations, as well as analytical results have been obtained on the complete graph \cite{Nyc:Szn:Cis:12} and on various random graphs \cite{Jed:17}. Still, heuristically it was not explained why discontinuous phase transition appears in such a model.

Before we proceed to the heuristic explanation of the observed phenomena, \textbf{let us summarize the most important results}. Within the model with independence: (1) a discontinuous phase transition is observed, (2) from the completely ordered initial state the jump of the order takes place at $p^*=2(1-r)$ and the jump size is equal $r-1/2$, (3) consequently for $r=0.5$ the transition is continuous. Within the model with anticonformity: (1) for $p<p^*=1-r$ the stationary concentration of the positive opinion  decays with $p$ and for $p>p^*=1-r$ it has a constant value $c_{st}=r$ or $c_{st}=1-r$ depending on the initial value of $c(0)$, (2) consequently for $r=0.5$ a continuous phase transition is observed at $p^*=0.5$.

\subsection*{Heuristic explanation of the obtained results}
We start with the \textbf{heuristic explanation for independence}. If the fraction of up spins is smaller than threshold $r$, i.e. $c(t)<r$ at any time $t$ and simultaneously the fraction of down spins is smaller than threshold $r$ , i.e. $1-c(t)<r \rightarrow c(t)>1-r$ then, due to the model's assumptions, conformity cannot take place and changes are caused exclusively by independence. Therefore, for $c(0) \in (1-r,r)$ there are only random flips in both directions and in result we should obtain stationary value $c_{st}=0.5$, independently on $p$. Indeed we obtain such a result from simulations, as well as from analytical reasoning. For $c(0)>r$ only up spins can cause conformity and therefore: with probability $p$ spins flip randomly and with probability $1-p$ they flip up. This means that for $p=0$ the system will always reach ordered state with all spins up and thus $c_{st}=1$. On the other hand, for $p=1$ only random flips will occur, which leads to  $c_{st}=1/2$. Expecting the linear dependence between $c_{st}$ and $p$ we can immediately draw the line, which crosses these two points $(p,c_{st})=(0,1)$ and $(p,c_{st})=(1,1/2)$, i.e. $c_{st}=1-p/2$. Analogous reasoning can be provided for $c(0)<1-r$. In such a case the line $c_{st}=c_{st}(p)$ should cross points $(p,c_{st})=(0,-1)$ and $(p,c_{st})=(1,1/2)$, which gives $c_{st}=3p/2-1$.
Alternative way to deduce the above relation is to realize that in general a fraction of $p$ spins flips with probability $1/2$ competing which the complete order $c_{st}=1$, if only $c(0)>r$ and thus:
\begin{eqnarray}
\begin{array}{l}
c_{st}  =
\left\{
\begin{array}{ll}
1-\frac{1}{2}p & \mbox{for} \hspace{0.2cm} p \le p^{*}  \\
\frac{1}{2} & \mbox{for} \hspace{0.2cm} p>p^{*} .\\
\end{array}
\right.
\end{array}
\label{eq:cst_indep}
\end{eqnarray}
We obtain the value of $p^{*}$ from the condition:
\begin{equation}
r=1-\frac{1}{2}p^* \rightarrow p^{*}=2(1-r),
\end{equation}
because, as written above, for $c(0)<r$ the system will reach disorder, i.e. $c=1/2$. It should become clearer looking at Fig. \ref{fig:flow}. Exactly the same results were obtained from the Monte Carlo simulations, as well as the mean-field approach. However, this simple explanation helps to realize that the jump of $c_{st}$ at $p^*=2(1-r)$ is caused just because conformity cannot take place if $c(t) \in (1-r,r)$ and thus in this range the system is attracted to the disordered state with up-down symmetry.

\textbf{For the model with anticonformity} heuristic explanation is equally simple. If the fraction of up spins is smaller than threshold $r$, i.e. $c(0)<r$ and simultaneously the fraction of down spins is smaller than threshold $r$ , i.e. $1-c(0)<r \rightarrow c(0)>1-r$ then there is not enough social pressure for conformity, as well as for anticonformity. Therefore, for $c \in (1-r,r)$ there are no changes in the system and $\forall_t c(t)=c(0)$. For $c(0)>r$ only up spins drive changes in the system: with probability $p$ a spin flips from $\uparrow$ to $\downarrow$ or with probability $1-p$ a spin flips from $\downarrow$ to $\uparrow$, and thus
\begin{eqnarray}
\begin{array}{l}
c  =
\left\{
\begin{array}{ll}
1-p & \mbox{for} \hspace{0.2cm} p \le p^* \\
1-p^* & \mbox{for} \hspace{0.2cm} p > p^*,
\end{array}
\right.
\end{array}
\label{eq:cst_anti}
\end{eqnarray}
where $p^*$ can be obtained analogously like for independence:
\begin{equation}
r=1-p^* \rightarrow p^{*}=1-r,
\end{equation}
Analogous reasoning we can perform for $c(0)<1-r$, the only difference is that then only down spins drives changes in the system, so again the results will be symmetrical with respect to the line $c_{st}=1/2$.

\section*{Discussion}
The modification of the Watts threshold model  that we have proposed here maybe treated as a destruction of the model from the social point of view. However, our aim was not to propose a model describing properly some social phenomena but to understand the nature of the phase transitions observed within models of binary opinions with a single-flip dynamics and up-down symmetry.

It should be noticed that the Model A, proposed here, can be treated as the generalization of the original majority-vote model, which corresponds to $r=1/2$. Within such a model only continuous phase transitions are observed, even in the presence of an additional noise \cite{Vie:Cro:16,Enc:etal:19}. In \cite{Vie:Cro:16} the model was studied via the mean-field approximation on the square lattice and it was suggested that maybe for larger number of neighbors discontinuous phase transitions would be observed. However, the same model was recently examined on various graphs and it was shown that the presence of independence does not change the type of the phase transition,
which remains continuous even for highly connected networks \cite{Enc:etal:19}. Indeed we show here that for $r=1/2$ transition is always continuous even in the limiting (the most connected) case of a complete graph. However, we have shown that discontinuous phase transition may appear within the model with independence for $r>1/2$. This result agrees with those obtained for the threshold $q$-voter model, for which also discontinuous phase transition is possible only if $r>1/2$ \cite{Nyc:Szn:13}.

However, results obtained here help not only to understand the difference between the $q$-voter model and the majority-vote model. Additionally, they help to understand what is a difference between anti-conformity and independence. For both variants of the model proposed here, conformity cannot take place for $c(t) \in (1-r,r)$ and thus ordering cannot occur. However, there is big difference between Model I and Model A, that cannot be visible for $r=1/2$: independence can take place for $c(t) \in (1-r,r)$, whereas anticonformity cannot. In result for Model I there is an attracting point $c_{st}=1/2$ for any value of $p$, whereas freezing of the system, i.e. $\forall_t c(t)=c(0)$ within Model A. This causes jump between partially ordered state to the disordered state within Model I, whereas freezing (and in consequence a lack of jump) within Model A.

Threshold $r>1/2$ plays one more role, namely introduces a kind inertia on the microscopic level. For $r=1/2$ a spin always takes a position of majority, independently of its state, whereas for $r>1/2$ it may happen that there is no majority above the threshold $r$ and in such a case neighborhood does not influence a target spin. This point of view is particularly interesting if we recall recent results obtained in \cite{Che:etal:17,Enc:etal:18}. It has been shown that  inertia introduced on the microscopic level into the majority-vote model, which causes that spin-flip probability of a given spin depends not only on the states of its neighbors, but also on its own state, can change the type of the phase transition from continuous to discontinuous. Interestingly, in \cite{Enc:etal:18} the dependence between the order parameter, i.e. average magnetization $<m>$,  and the control parameter, i.e. the probability of anticonformity (denoted by $f$), was also nearly linear for $f<f^*$. Moreover $<m>=0$ was stable for $f \in (0,f^*)$, similarly as in our model.

Another possibility to obtain a discontinuous phase transition has been recently suggested within the generalized threshold $q$-voter model, in which two types of non-conformity were introduced simultaneously \cite{Nyc:etal:18}. It has been shown that discontinuous phase transition could be obtained even without independence if only the threshold for anticonformity  would be smaller than for conformity. In the face of the reasoning carried out here, this result also becomes obvious, because than the "forbidden" range for conformity is larger than for anticonformity and the jump will increase with the difference between these two thresholds.

We are aware of the fact that many people may wonder why to care about the type of the phase transition. Is there any reason, other than academic, to distinguish between continuous and discontinuous phase transitions? In the face of the social observations, and more recently also laboratory experiments, it seems that discontinuous phase transitions are particularly important, mainly because of the notion of the social hysteresis and the critical mass \cite{Bee:Sum:Rat:01,Sche:Wes:Bro:03,Pru:etal:18,Doe:etal:18,Cen:etal:18}. Both phenomena are strictly related to discontinuous phase transitions. There is no hysteresis within continuous phase transitions, which means that the state of the system is fully determined by the external conditions and does not depend on the history of the system. On the other hand, hysteresis, which is observed within discontinuous phase transitions, means that under the same external conditions the system can be in a different states depending on its previous states (history). The social hysteresis was observed in animal \cite{Bee:Sum:Rat:01,Pru:etal:18,Doe:etal:18} as well as in human societies \cite{Els:76,Sche:Wes:Bro:03,Cla:03}. The second phenomena, so-called critical mass, which was recently observed experimentally in social convention \cite{Cen:etal:18} is also strongly related to the discontinuous phase transitions. Within the continuous phase transitions, there is no phase coexistence and the transition appears strictly at a given critical point due to the fluctuations and the infinite range of the correlations between them. Because of these correlations, the transition takes place immediately and simultaneously in the entire system. On the other hand, a seed  (a "critical mass") initiating the transition is needed to change the phase under discontinuous phase transition.

We realize that more interesting results could be obtained for different homogeneous and heterogeneous networks. Moreover, both noises could be introduced simultaneously as in \cite{Vie:Cro:16,Enc:etal:19,Nyc:etal:18}. However, the aim of this work was different --  we wanted to understand the reason for the discontinuous phase transition in the simplest possible settings. Still, we believe that further studies of the model proposed here would be desirable task for the future, since it is a simple generalization of the majority-vote rule.

\section*{Acknowledgements}
This work was supported by funds from the National Science Center (NCN, Poland) through grant no. 2016/21/B/HS6/01256.


%
%
%
%
\end{document}